\documentclass[12pt]{article}
\usepackage{amsmath}
\usepackage{graphicx}
\usepackage{url}
\usepackage{geometry}
\usepackage{caption}
\usepackage{placeins}
\usepackage{setspace}
\usepackage{comment}
\usepackage{ragged2e}
\usepackage{rotating}
\usepackage{array}
\usepackage{lscape}
\usepackage{float}
\usepackage{hyperref}
\usepackage{tikz}
\usepackage{lscape}
\usepackage{siunitx}
\usetikzlibrary{positioning}
\usepackage[superscript]{cite}
\newcommand{\blind}{1}

\begin{document}

\def\spacingset#1{\renewcommand{\baselinestretch}%
{#1}\small\normalsize} \spacingset{1}

\if1\blind
{
  \title{\bf Estimating Longitudinal Causal Effects with Unobserved Noncompliance Using a Semi-Parametric G-computation Algorithm}
  \author{Ross L Peterson, David M Vock, and Joseph S Koopmeiners\hspace{.2cm}\\
    Division of Biostatistics, School of Public Health, University of Minnesota\\}
  \maketitle
} \fi

\if0\blind
{
  \bigskip
  \bigskip
  \bigskip
  \begin{center}
    {\LARGE\bf Title}
\end{center}
  \medskip
} \fi

\bigskip
\begin{abstract}
Participant noncompliance, in which participants do not follow their assigned treatment protocol, often obscures the causal relationship between treatment and treatment effect in randomized trials. In the longitudinal setting, the G-computation algorithm can adjust for confounding to estimate causal effects. Typically, G-computation assumes that both 1) compliance is observed; and 2) the densities of the confounders can be correctly specified. We aim to develop a G-computation estimator in the setting where both assumptions are violated. For 1), in place of unobserved compliance, we substitute in probability weights derived from modeling a biomarker associated with compliance. For 2), we fit semiparametric models using predictive mean matching. Specifically, we parametrically specify only the conditional mean of the confounders, and then use predictive mean matching to randomly generate confounder data for G-computation. In both the simulation and application, we compare multiple causal estimators already established in the literature with those derived from our method. For the simulation, we generated data across different sample sizes and levels of confounding. For the application, we apply our method to a trial that sought to evaluate the effect of cigarettes with low nicotine on cigarette consumption (Center for the Evaluation of Nicotine in Cigarettes Project 2 - CENIC-P2).
\end{abstract}

\vfill

\newpage
\spacingset{1.5}

\section{Introduction}

Randomized controlled trials (RCTs) are viewed as the gold standard for testing the efficacy of new treatments in target populations of interest. To establish causation due to the treatment, RCTs depend on properly randomizing participants to different treatment arms. In some trials, however, participants are not obligated to comply with their randomized treatment assignment. Noncompliance most often occurs when participants must self-administer the treatment without the supervision of study personnel or a medical practitioner.\cite{stephenson93} Participants may then choose to either comply or deviate from the treatment protocol for a variety of reasons (e.g., due to advent of side effects, insufficient benefit, or availability of commercial alternatives to the treatment).\cite{boza87} Additionally, in longitudinal studies, multiple time points present multiple opportunities for participants to be noncompliant. 

When participants do not comply, the treatment received may be different from the treatment randomly assigned. Primary analyses of such trials must then be qualified with the consideration that not all participants complied with the trial. Specifically, both the intention-to-treat (ITT) estimate and the Per Protocol estimate based on self-reported compliance may not match the treatment effect if all participants had complied (i.e., the causal effect).\cite{bellamy07} For the ITT estimate, noncompliant participants do not receive the full dose of treatment and as such may have poor study outcomes. This may dilute the ITT estimate.\cite{van07} For the Per Protocol estimate, participants who self-reported compliance may systematically differ on some confounding variable.\cite{jin08}

For treatments that could be mandated by federal law to effectively force compliance, the causal effect is more relevant than the ITT estimator.\cite{koopmeiners19} As a motivating example, in the United States, the Family Smoking Prevention and Tobacco Control Act provides the Food and Drug Administration with the regulatory authority to reduce (but not eliminate) the nicotine content in commercial cigarettes if it would improve public health. Regulatory tobacco trials seek to evaluate the effect of nicotine reduction on measures of smoking behavior. In particular, we were motivated by the Center for the Evaluation of Nicotine in Cigarettes Project 2 (CENIC-P2) trial, a 20-week RCT that investigated the effect of very low nicotine content (VLNC, with 0.4mg of nicotine per gram of tobacco) cigarettes versus normal nicotine content (NNC, with 15.5mg of nicotine per gram of tobacco) cigarettes on average number of cigarettes smoked per day. During the follow-up period, participants were asked to only smoke their assigned study cigarettes but could additionally smoke commercial cigarettes (i.e., non-study cigarettes) with normal nicotine content.

In longitudinal studies, as noncompliance, the study outcome, and the confounding variables can all vary over time, there may be time-varying confounding. In the presence of both time-varying noncompliance and time-varying confounding, standard regression-based methods cannot be used to estimate the causal effect.\cite{rosenbaum83} Longitudinal causal estimators that can consistently estimate the causal effect of an intervention with time-varying noncompliance include the G-computation algorithm, inverse probability of compliance weighted (IPCW) estimators, and structural nested mean models estimated with G-estimation.\cite{robins86,robins97,hernan06,cain09,robins89,robins92}

For CENIC-P2, both IPCW estimators and G-estimation may be less than desirable. Specifically, for IPCW estimators, the denominator of the weights of compliance is a product of the compliance probabilities at each time point. As there are many time points for compliance in CENIC-P2 (i.e., 5 time points), the product of many compliance probabilities could create bias and variability in estimation of the weights and hence estimation of the causal effect.\cite{robins08} Additionally, a primary advantage of G-estimation is that structural nested mean models can directly model the interactions between time-varying treatments and time-varying confounding.\cite{naimi17} This can help identify the optimal treatment regime; however, in CENIC-P2, we are only interested in modeling a single treatment regime (i.e., compliance across all time points of the trial). Thus, there are no time-varying treatments in the model fit for the study outcome, and hence no need to model any interactions that include them. 

We thus turn our attention to the G-computation algorithm. To adjust for time-varying confounding, G-computation models both the time-varying confounders and the study outcome conditional on the treatment and confounder histories. The causal effect is then estimated by randomly sampling from the fitted densities.

Like all other longitudinal causal estimators, G-computation assumes that compliance is observed. However, in trials where participants must self-administer the treatment, compliance is not directly observed. Various methods have been devised to determine compliance when unobserved (e.g., self-reports), but most all depend on subjective or indirect information, and as such are unreliable.\cite{farmer99,dimatteo04}

In some trials, participants' biomarkers may systematically change in response to the treatment or any alternatives to provide objective information about noncompliance. Consider three recently published regulatory tobacco clinical trials studying the effect of VLNC cigarettes versus NNC cigarettes on smoking behavior.\cite{donny15,hatsukami18,smith19} At each study visit for each trial, investigators measured participants' biomarkers of nicotine exposure (e.g., total nicotine equivalents (TNE), which measures most nicotine metabolites in the urine) which can be used to detect noncompliance to VLNC cigarettes.\cite{benowitz15,denlinger16,boatman19} Without having to observe compliance, Boatman et al.~(2017) used TNEs to re-weight an IPCW estimator to estimate the causal effect of VLNC cigarettes on average number of cigarettes smoked per day.\cite{boatman17} However, Boatman et al.'s method is limited to examining information collected at a single time point.

We will develop a G-computation estimator that uses biomarker information in place of unobserved compliance. The motivating dataset from CENIC-P2, however, presents one additional challenge for G-computation. The confounders include survey instruments whose supports are bounded with a number of observations near the ends of the scales. This makes it difficult to correctly specify parametric conditional densities. While we could fit fully parametric regression models, incorrectly specifying the conditional densities may lead to biased estimators. Instead, we will implement predictive mean matching, which requires that we only correctly specify the conditional mean as opposed to the full conditional distribution.

The goal of this research is thus two-fold: to develop a G-computation estimator that can 1) account for unobserved noncompliance; and 2) semiparametrically model confounders with densities that may be difficult to parametrically specify. We first evaluate our method by simulation, investigating the effects of both different levels of confounding and sample sizes on estimation of the causal mean. We also apply our method to the data collected in CENIC-P2.

\section{Methods}

\subsection{Dataset and Target of Inference}

We consider a longitudinal RCT, where $i \in \{1,\ldots,n\}$ denotes the participant, $j \in \{0,\ldots,K\}$ denotes the time point, and $A_i = a_i$ denotes the randomized treatment group of participant $i$. For our purposes, we treat the true compliance status as unobserved and binary; that is, each participant can either fully comply or not fully comply at each time point. Let $C_{ij}$ be the true, unobserved compliance indicator ($1$ denotes compliance, $0$ denotes noncompliance) for the $i$th participant at the $j$th time point for $j \in \{1,\ldots,K\}$. As participant compliance can vary over time, let $\bar{c}_{iK}$ be one of the $2^K$ compliance patterns for $K$ time points for participant $i$. For example, for $K=2$ time points, the set of compliance patterns has length $2^2 = 4$ and consists of: $\{\{0,0\},\{1,0\},\{0,1\},\{1,1\}\}$. We use the overbar notation to denote history (e.g., for full compliance, $\bar{c}_{iK} = \{1,\ldots,1\}$).

Define $Y_{iK}^*(a_i,\bar{c}_{iK})$ to be the study outcome at the end of the trial of a randomly selected participant if, possibly contrary to fact, we set $A_i = a_i$ and $\bar{C}_{iK} = \bar{c}_{iK}$. Because for each participant we do not observe $Y_{iK}^*(a_i,\bar{c}_{iK})$ for all $a_i$ and $\bar{c}_{iK}$, $Y_{iK}^*(a_i,\bar{c}_{iK})$ is a potential outcome. The target of inference is the expected difference in the outcome among randomized treatment groups $a_i$ and $a_i'$ if all participants were to be compliant. That is:
\begin{equation*}
E(Y_{iK}^*(a_i,\bar{c}_{iK}=\{1,\ldots,1\}) - Y_{iK}^*(a_i',\bar{c}_{iK}=\{1,\ldots,1\})).
\end{equation*}
Define $L_{ij} = \{Y_{ij},Z_{ij}\}$ as the set of time-varying confounders for $j \in \{0,\ldots,K\}$ with history $\bar{L}_{ij} = \{L_{i0}\ldots,L_{ij}\}$, where $Y_{ij}$ is the study outcome measured at both the end of the trial and at earlier time points and $Z_{ij}$ is an additional set of confounders. Let $W_i = \{L_{i1}^*(a_i,c_{i1}),\ldots,L_{iK}^*(a_i,\bar{c}_{iK})\}$ be the set of potential outcomes for $L_{ij}$ for the $i$th participant across all $a_i$ and $\{c_{i1},\ldots,\bar{c}_{iK}\}$. Additionally, let $B_{ij}$ be the biomarker for $j \in \{1,\ldots,K\}$, let $X_i$ be the set of time-invariant confounders, and $D_{ij}$ be the self-reported indicator of compliance ($1$ denotes self-reported compliance, $0$ denotes self-reported noncompliance) for $j \in \{1,\ldots,K\}$.

\subsection{Identifying Assumptions}

To estimate the expected value of the potential outcome based on the observed data, we make the following identifying assumptions.\cite{robins08} First, we make the no unmeasured confounders assumption, where we assume that compliance status is only confounded by $L_{ij}$ and $X_i$. Given this information, compliance status at each time point does not depend on the potential outcomes (i.e., $P(C_{ij}=1|\bar{L}_{ij},X_i,A_i=a_i,W_i) = P(C_{ij}=1|\bar{L}_{ij},X_i,A_i=a_i)$). Second, we make the positivity assumption, where we assume that there is a non-zero probability of compliance at each time point across all values of the confounders (i.e., $P(C_{ij}=1|\bar{L}_{ij},X_i,A_i=a_i) > 0$). Third, we make the consistency assumption, where we assume that the mode of compliance is immaterial. That is, at each time point, a participant who voluntarily complies has the same outcome relative to if he or she were forced to be compliant (i.e., if $A_i = a_i$ and $\bar{C}_{ij} = \bar{c}_{ij}$, then $Y_{ij} = Y_{ij}^*(a_i,\bar{c}_j)$).

\subsection{The G-computation Algorithm}

We delineate the G-computation algorithm as follows. Under the identifying assumptions, the joint density of both the potential outcomes under full compliance (i.e., $\{L_{i1}^*(a_i,c_{i1}=1),\ldots,L_{iK}^*(a_i,\bar{c}_{iK}=\{1,\ldots,1\})\}$) and $X_i$ is equal to:
\begin{equation} \label{eq:a}
\begin{split}
&f(L_{iK}|\bar{L}_{iK-1},X_i,A_i=a_i,\bar{C}_{iK}=\{1,\ldots,1\})\\&* f(L_{iK-1}|\bar{L}_{iK-2},X_i,A_i=a_i,\bar{C}_{iK-1}=\{1,\ldots,1\}),\ldots, f(L_{i0}|X_i)f(X_i),
\end{split}
\end{equation}
where $f(L_{iK}|\bar{L}_{iK-1},X_i,A_i=a_i,\bar{C}_{iK}=\{1,\ldots,1\})$ is the conditional density of $L_{iK}$ given $\bar{L}_{iK-1},X_i,A_i=a_i,\bar{C}_{iK}=\{1,\ldots,1\}$. Then the density of $Y_{iK}^*(a_i,\bar{c}_{iK}=\{1,\ldots,1\})$ is given by:
\begin{equation} \label{eq:m}
\begin{split}
\int \ldots \int &f(L_{iK}|\bar{L}_{iK-1},X_i,A_i=a_i,\bar{C}_{iK}=\{1,\ldots,1\})f(L_{iK-1}|\bar{L}_{iK-2},X_i,A_i=a_i,\bar{C}_{iK-1}=\{1,\ldots,1\})\\&,\ldots, f(L_{i0}|X_i)f(X_i) dX_i dL_{i0} \ldots dL_{iK-1} dZ_{iK}.
\end{split}
\end{equation}
We can estimate $E(Y_{iK}^*(a_i,\bar{c}_{iK}=\{1,\ldots,1\}))$ with three steps: 
\begin{enumerate}
\item We model the densities of the baseline confounders, $L_{i0}$ and $X_i$, often with the empirical distributions.
\item For each subsequent time point, we model the conditional confounder density of $L_{ij}$ given past covariate history, often with parametric models.
\begin{itemize}
\item[--] Both the estimated densities for steps 1 and 2 can then be plugged into (\ref{eq:a}).
\end{itemize}
\item We use Monte Carlo integration to approximate the integrals in (\ref{eq:m}) to estimate the distribution of $Y_{iK}^*(a_i,\bar{c}_{iK}=\{1,\ldots,1\})$ as well as any corresponding summary statistics (e.g., the mean).
\end{enumerate}

We discuss steps 2 and 3 in detail. In the setting where the compliance indicator $C_{ij}$ is known, we could assume a parametric model for the density of $L_{ij}$ indexed by parameter vector $\beta_j$ for time point $j$. We could then estimate $\beta_j$ by solving the following score equations:
\begin{equation} \label{eq:full}
\sum\limits_{i=1}^n I(\bar{C}_{ij}=\{1,\ldots,1\}) \frac{\partial}{\partial \beta_j} \log f(L_{ij}|\bar{L}_{ij-1},X_i,A_i=a_i,\bar{C}_{ij}=\{1,\ldots,1\};\beta_j)=0.
\end{equation}
In many applications, we may be willing to assume that the conditional density of $L_{ij}$ given $\bar{L}_{ij-1},X_i,A_i=a_i,\bar{C}_{ij}=\{1,\ldots,1\}$ equals the conditional density of $L_{ij}$ given $L_{ij-1},X_i,A_i=a_i,C_{ij}=1$ That is, the conditional density only depends on the most recent $L_{ij-1}$ and compliance at the current time point, in addition to the time-invariant confounders and treatment arm. In that case, the score equations in (\ref{eq:full}) simplify to:
\begin{equation*}
\sum\limits_{i=1}^n I(C_{ij}=1) \frac{\partial}{\partial \beta_j} \log f(L_{ij}|L_{ij-1},X_i,A_i=a_i,C_{ij}=1;\beta_j)=0.
\end{equation*}
Note that in some contexts we may be willing to assume that $\beta_1 = \ldots = \beta_K = \beta$ (i.e., the parameters are shared across time points). We could then estimate $\beta$ by solving the following estimating equations:
\begin{equation} \label{eq:s}
\sum \limits_{j=1}^K \sum\limits_{i=1}^n I(C_{ij}=1) \frac{\partial}{\partial \beta} \log f(L_{ij}|L_{ij-1},X_i,A_i=a_i,C_{ij}=1;\beta)=0.
\end{equation}
As $L_{ij} = \{Y_{ij},Z_{ij}\}$, it may be difficult to specify a multidimensional joint density. Instead, we can model the conditional density of $L_{ij}$ by factorizing $f(L_{ij}|L_{ij-1},X_i,A_i=a_i,C_{ij}=1;\beta)$ into individual components for $Z_{ij}$ given $L_{ij-1},X_i,A_i=a_i,C_{ij}=1$ and $Y_{ij}$ given $Z_{ij},L_{ij-1},X_i,A_i=a_i,C_{ij}=1$. Then:
\begin{equation} \label{eq:f}
\begin{split}
&f(L_{ij}|L_{ij-1},X_i,A_i=a_i,C_{ij}=1;\beta) \\&=h(Y_{ij}|Z_{ij},L_{ij-1},X_i,A_i=a_i,C_{ij}=1;\beta_Y)s(Z_{ij}|L_{ij-1},X_i,A_i=a_i,C_{ij}=1;\beta_Z),
\end{split}
\end{equation}
where parameter vectors $\beta_Y$ and $\beta_Z$ correspond to the subsets of $\beta$ for the conditional densities of $Y_{ij}$ and $Z_{ij}$, respectively. We can further factorize the conditional density of $Z_{ij}$ when $Z_{ij}$ is multivariate.

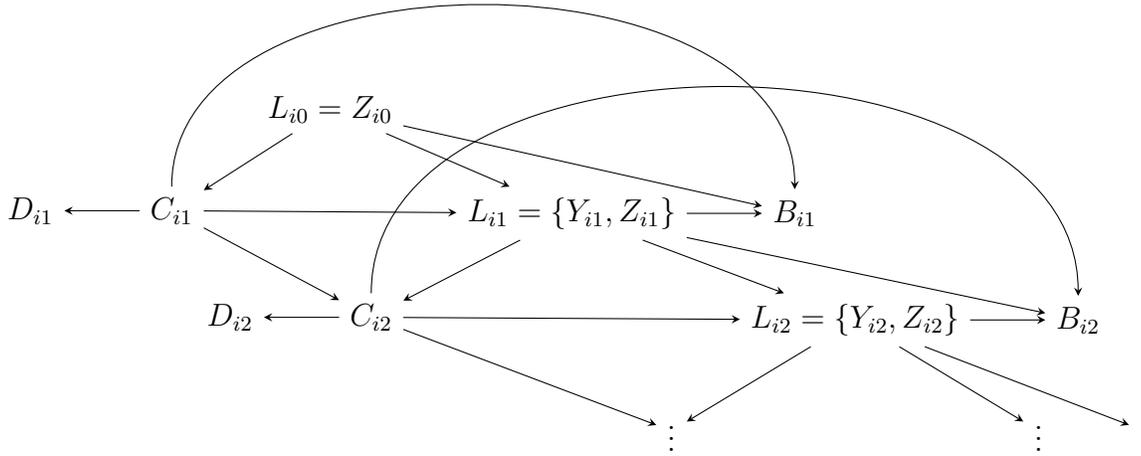
\begin{figure}[H]
\caption{One plausible directed acyclic graph for CENIC-P2. The time-invariant confounders $X_i$ affect all $L_{ij}$, $C_{ij}$ and $B_{ij}$, but not $D_{ij}$.}
\begin{tikzpicture}[%
	->,
	>=stealth,
	node distance=1cm,
	pil/.style={
		->,
		thick,
		shorten =2pt,}
	]
	\node (1) {$L_{i0} = Z_{i0}$};
	\node[below left = of 1] (2) {$C_{i1}$};
	\node[below right = of 1] (3) {$L_{i1} = \{Y_{i1},Z_{i1}\}$};
	\node[right = of 3] (4) {$B_{i1}$};
	\node[below left = of 3] (5) {$C_{i2}$};
	\node[below right = of 3] (6) {$L_{i2} = \{Y_{i2},Z_{i2}\}$};
    \node[right = of 6] (7) {$B_{i2}$};
	\node[below left = of 6] (8) {$\vdots$};
	\node[below right = of 6] (9) {$\vdots$};
	\node[left = of 2] (10) {$D_{i1}$};
	\node[left = of 5] (11) {$D_{i2}$};
	\node[right = of 9] (12) {$\vdots$};
	\draw [->] (1) to (2);
	\draw [->] (1) to (3);
	\draw [->] (2) to (3);
	\draw [->] (3) to (4);
	\draw [->] (2) to (5);
	\draw [->] (1) to (4);
	\draw [->] (2) to [out=90,in=90] (4);
	\draw [->] (3) to (5);
	\draw [->] (3) to (6);
	\draw [->] (5) to (6);
	\draw [->] (6) to (8);
	\draw [->] (6) to (9);
	\draw [->] (6) to (7);
	\draw [->] (5) to (8);
	\draw [->] (3) to (7);
	\draw [->] (5) to [out=90,in=90] (7);
	\draw [->] (2) to (10);
	\draw [->] (5) to (11);
	\draw [->] (6) to (12);
\end{tikzpicture}
\label{fig:dag}
\end{figure}

To implement Monte Carlo integration, we use the following steps for random samples $r \in \{1,\ldots,R\}$. First, we randomly sample both $l_{i0,r}$ and $x_{i,r}$ from the empirical distributions of $L_{i0}$ and $X_i$. We then randomly sample $l_{i1,r}$ from the fitted density $f(L_{i1}|L_{i0}=l_{i0,r},X_i=x_{i,r},A_i=a_i,C_{i1}=1;\hat{\beta})$. For each subsequent time point, we randomly sample from $f(L_{ij}|L_{ij-1}=l_{ij-1,r},X_i=x_{i,r},A_i=a_i,C_{i1}=1;\hat{\beta})$. The average of the randomly sampled $y_{iK,r}$ at the last time point is the causal estimator:
\begin{equation*}
\hat{E}(Y_{iK}^*(a_i,\bar{c}_{iK}=\{1,\ldots,1\})) = \frac{1}{R} \sum\limits_{r = 1}^R y_{iK,r}.
\end{equation*}
Given that we assume that observations only depend on data collected at the current and previous time points, Figure \ref{fig:dag} details one plausible directed acyclic graph (DAG) for how the data might have been generated in CENIC-P2. The DAG assumes that self-reported compliance $D_{ij}$ is conditionally independent of all other variables given compliance $C_{ij}$.
\captionsetup[figure]{labelfont=bf,font=normalsize,justification=raggedright,singlelinecheck=false}

\subsection{The G-computation Algorithm with Unobserved Participant Noncompliance}

CENIC-P2 presents two challenges for G-computation: 1) compliance is unknown; and 2) the confounders come from densities that may be difficult to parametrically specify. To resolve 1), we can use both the observed biomarker $B_{ij}$ that systematically responds to compliance $C_{ij}$ and self-reported compliance $D_{ij}$. We insert probabilities of compliance conditional on the biomarker, self-reported compliance, time-varying confounders (measured at the current and previous time points), time-invariant confounders, and treatment arm in place of the indicator of compliance in (\ref{eq:s}):
\begin{equation} \label{eq:wlog}
\sum \limits_{j=1}^K \sum\limits_{i=1}^n P(C_{ij}=1|L_{ij},L_{ij-1},B_{ij},X_i,D_{ij},A_i=a_i) \log \frac{\partial}{\partial \beta} f(L_{ij}|L_{ij-1},X_i,A_i=a_i,C_{ij}=1;\beta)=0.
\end{equation}
Moreover, by iterated expectation, the expected value of the estimating function in (\ref{eq:s}) is equal to the expected value of (\ref{eq:wlog}) (see Supplementary Material section \ref{ie}). Thus, (\ref{eq:wlog}) is also a mean-zero estimating function. Therefore, under suitable regularity conditions, estimators of $\beta$ from solving (\ref{eq:wlog}) are consistent and asymptotically normal.\cite{stefanski02}

Additionally, in most trials there is usually no incentive to report noncompliance. Therefore, we typically assume that self-reports of noncompliance are accurate, where $D_{ij} = 0$ implies $C_{ij} = 0$ (i.e., $P(C_{ij}=1|L_{ij},L_{ij-1},B_{ij},X_i,D_{ij}=0,A_i=a_i)=0$). Observations with $D_{ij} = 0$ thus do not contribute to parameter estimation.

Although the probability of compliance (i.e., $P(C_{ij}=1|L_{ij},L_{ij-1},B_{ij},X_i,D_{ij},A_i=a_i)$) is unknown and cannot be directly estimated from the data (e.g., by specifying a logistic regression model with $C_{ij}$ as the response variable), we can derive a consistent estimator as a function of directly estimable densities. By Bayes' rule, we can write the probability of compliance as a function of the probability without the biomarker and the biomarker conditional on compliance. Specifically, $P(C_{ij}=1|L_{ij}=l_{ij},L_{ij-1}=l_{ij-1},B_{ij}=b_{ij},X_i=x_i,D_{ij},A_i=a_i)$ can be expressed as:
\begin{equation*}
\frac{\rho(l_{ij},l_{ij-1},x_i,a_i,d_{ij};\alpha)g(b_{ij}|l_{ij},l_{ij-1},x_i,a_i,d_{ij},c_{ij}=1;\xi)}{\psi(b_{ij}|l_{ij},l_{ij-1},x_i,a_i,d_{ij};\alpha,\xi)},
\end{equation*}
where $\rho(l_{ij},l_{ij-1},x_i,a_i,d_{ij};\alpha) = P(C_{ij}=1|L_{ij}=l_{ij},L_{ij-1}=l_{ij-1},X_i=x_i,A_i=a_i,D_{ij};\alpha)$ indexed by parameter vector $\alpha$, $g$ is the conditional density of $B_{ij}$ indexed by parameter vector $\xi$, and the denominator is equal to:
\begin{equation} \label{eq:mix}
\begin{split}
&\psi(b_{ij}|l_{ij},l_{ij-1},x_i,a_i,d_{ij};\alpha,\xi) \\&= \rho(l_{ij},l_{ij-1},x_i,a_i,d_{ij};\alpha)g(b_{ij}|l_{ij},l_{ij-1},x_i,a_i,d_{ij},c_{ij}=1,\xi) \\ &+ (1-\rho(l_{ij},l_{ij-1},x_i,a_i,d_{ij};\alpha))g(b_{ij}|l_{ij},l_{ij-1},x_i,a_i,d_{ij},c_{ij}=0;\xi).
\end{split}
\end{equation}
As all variables but $C_{ij}$ are observed, we can directly estimate the components of $P(C_{ij}=1|L_{ij},L_{ij-1},B_{ij},X_i,A_i=a_i,D_{ij})$ from the mixture density in (\ref{eq:mix}). As is often the case with mixture densities such as (\ref{eq:mix}), it would be difficult to directly solve for the maximum likelihood estimates. Instead, we can use the EM algorithm.\cite{dempster77} Supplementary Material section \ref{ca} delineates the E-step and M-step.

\subsection{The G-computation Algorithm with Predictive Mean Matching}

As it may not be reasonable to fit parametric models to the densities of the confounders in (\ref{eq:f}), we instead fit semiparametric models using predictive mean matching.\cite{rubin86,little88} For now, assume that the conditional density of $Y_{ij}$ can be correctly specified using parametric models while the conditional density of $Z_{ij}$ may not be correctly specified.

While we may not be willing to assume a fully parametric density (e.g., a normal distribution) for $s(Z_{ij}|L_{ij-1},X_i,A_i=a_i,C_{ij}=1;\beta_Z)$, we can instead assume that only the conditional mean (i.e., $E(Z_{ij}|L_{ij-1},X_i,A_i=a_i,C_{ij}=1;\beta_Z)$) can be parametrically specified where $\beta_Z$ solely consists of regression coefficients. We can consistently estimate $\beta_Z$ by solving a mean-zero estimating function such as the following:
\begin{equation*}
\begin{split}
\sum\limits_{j=1}^K\sum\limits_{i=1}^n &P(C_{ij}=1|L_{ij},L_{ij-1},B_{ij},X_i,A_i=a_i,D_{ij})\\&*\{Z_{ij}-E(Z_{ij}|L_{ij-1},X_i,A_i=a_i,C_{ij}=1;\beta_Z)\} \frac{\partial}{\partial\beta_Z}E(Z_{ij}|L_{ij-1},X_i,A_i=a_i,  C_{ij}=1;\beta_Z)=0.
\end{split}
\end{equation*}
The residual distribution of $Z_{ij}$, however, may not be well-approximated by a parametric distribution. Specifically, for CENIC-P2 $Z_{ij}$ comprises survey data with bounded supports. This poses a problem when we have to randomly sample from the conditional $Z_{ij}$ distribution for Monte Carlo integration, as the generated $Z_{ij}$ must fall within the survey bounds. Predictive mean matching can bring the data generation closer to randomly sampling from the true distribution. In predictive mean matching, the residual distribution is uniquely determined by the conditional mean; provided that the regression equation for the conditional mean is properly specified, predictive mean matching is an asymptotically unbiased estimator of the true distribution.\cite{chen00}

We delineate predictive mean matching as follows. At each time point, we first calculate the predicted values of the generated $Z_{ij}$ data and match them with the predicted values of the observed $Z_{ij}$ data across all time points. Then, we impute one of the corresponding observed $Z_{ij}$. Specifically, let the dot symbol $\dot{}$ denote the predicted value for the observed data and the hat symbol $\hat{}$ denote the predicted value for the generated data. For random samples $r \in \{1,\ldots,R\}$ with time points $j \in \{0,\ldots,K\}$, we can generate confounder data with the following steps:
\begin{enumerate}
\item Randomly sample baseline confounders $l_{i0,r}$ and $x_{i,r}$ from the empirical distributions of $L_{i0}$ and $X_i$, respectively.
\item For each subsequent time point, calculate $\hat{Z}_{ij,r}$ from the fitted $\hat{E}(Z_{ij}|L_{ij-1}=l_{ij-1,r},X_i = x_{i,r},A_i=a_i,C_{ij}=1;\hat{\beta}_Z)$.
\item Find the top five candidates of $\dot{Z}_i$ across all time points for which $|\dot{Z}_i-\hat{Z}_{ij,r}|$ is minimal.
\item Randomly select one $\dot{Z}_i$ and impute its corresponding observed $z_i$ for $z_{ij,r}$.
\item Randomly sample $y_{ij,r}$ from the fitted $h(Y_{ij}|Z_{ij}=z_{ij,r},L_{ij-1}=l_{ij-1,r},X_i = x_{i,r},A_i=a_i,C_{ij}=1;\hat{\beta}_1)$.
\item This generates $l_{ij,r} = \{y_{ij,r},z_{ij,r}\}$.
\end{enumerate}
This approach can easily be generalized to the case where $Z_{ij}$ contains multiple confounders and the density of $Y_{ij}$ may not be correctly specified.

Randomly sampling from some number of candidates according to proximity of predictive means is known as predictive mean random hot deck.\cite{van18} The selection criteria for matching is flexible; instead of the top five candidates, candidates could be selected based on some fixed threshold $\eta$ for $|\dot{Z}_i-\hat{Z}_{ij,r}| < \eta$ or the observed value for the top candidate could automatically be imputed.\cite{andridge10} We chose the top five candidates for two reasons: 1) relative to selecting candidates based on a fixed threshold $\eta$, the top five candidates ensure that there are always a limited number of candidates to draw from to expedite computation; and 2) relative to only selecting the top candidate, the top five candidates should provide more variability to protect against overfitting the empirical distributions.\cite{schenker96} 

Additionally, the CENIC-P2 data has two other characteristics which may make predictive mean matching more suitable. Specifically, predictive mean matching performs better for large sample sizes whose empirical distributions more closely approximate the true distributions relative to small sample sizes.\cite{lazzeroni90} Predictive mean matching also performs better when the data is not heavily skewed.\cite{kleinke17} For CENIC-P2, we have 1059 observations with confounders with densities moderately skewed towards the bounds.

\section{Simulation Study Design}

\subsection{Data Generation}

We used Monte Carlo simulation to assess our method assuming a single treatment arm $a_i$. Across different simulation scenarios, we varied both the sample size and the effect size of confounding on the study outcome $Y_{ij}$.

Across all scenarios, we assumed there to be single confounders for both $Z_{ij}$ and $X_i$. We generated longitudinal data in accordance with the DAG in Figure \ref{fig:dag}, where at each time point data is generated from other data collected at the current or previous time points. At baseline, $X_i$ is generated from a normal distribution while the study outcome $Y_{i0} = 0$. To mimic bounded confounder data, we generated $Z_{ij}$ from a zero-inflated poisson distribution. We then used linear regression to estimate the coefficients for the conditional mean of $Z_{ij}$, after which we used predictive mean matching to randomly sample the confounders in G-computation. Compliance $C_{ij}$ is generated from a bernoulli distribution with a logit link --- however, we were not able to generate compliance data in congruence with the density of compliance in (\ref{eq:mix}). The fitted logistic model is thus misspecified. We assumed that compliance varies across time points with a different intercept at each time point. Both $Y_{ij}$ and $B_{ij}$ were generated from normal distributions which can be correctly specified.

For the indicator of self-reported compliance at each time point, $D_{ij}$, we assumed that compliers honestly reported compliance and that noncompliers dishonestly reported compliance $\frac{2}{3}$ of the time (i.e., $D_{ij} = T^{1-C_{ij}}$, where $T \sim Bern(\frac{2}{3})$). Observations with $D_{ij} = 0$ did not contribute to parameter estimation.

We set the number of time points at $K = 6$ where $j \in \{0,\ldots,5\}$, analogous to CENIC-P2. We ran simulations for two sample sizes: $n = 500$ and $n = 1000$. Across all scenarios, data were generated such that over all participants and time points, approximately 40\% of observations were compliant while 60\% were noncompliant. After filtering out observations with self-reported noncompliance, approximately 80\% of observations equally distributed between compliance and noncompliance contributed to parameter estimation. Data were generated such that the percentage of compliance marginally increased with each time point.

The following statistics for the simulation only relate to self-reported compliers who we intend to analyze. To investigate the effect of confounding on estimation of the causal mean, we varied the size of $R^2$ in the model for the study outcome $Y_{ij}$ in (\ref{eq:f}). We simulated three values of $R^2$: 0.3, 0.5, and 0.7. Note that as confounding increases for $Y_{ij}$, the discrimination of the compliance probability in (\ref{eq:wlog}) increases. For $R^2 \in \{0.3,0.5,0.7\}$, the areas under the curve (AUC) of the compliance probabilities were $\{0.927,0.952,0.980\}$, respectively. The AUC values are within the range of those found in previous CENIC trials.\cite{boatman19} For $Z_{ij}$ over all time points, 8\% of observations were 0's while the mean of the poisson component was 33.

To evaluate each estimator, we calculated the empirical bias, Monte Carlo standard deviation (MC SD), and mean-squared error (MSE) across the estimates of the causal mean. Code to implement our method in the programming language R is available as a GitHub repository (\href{https://github.com/RPeterson4/Modified\_G\_computation}{https://github.com/RPeterson4/Modified\_G\_computation}).

\subsection{Proposed Estimators}

In both the simulation and application, we sought to compare a number of estimators of $E(Y_{iK}^*(a_i,\bar{c}_{iK}=\{1,\ldots,1\}))$. First, we have estimators already established in the literature, including the conventional ITT and Per Protocol estimators, the latter of which is based off of self-reported compliance. We also consider a regression-based estimator from Boatman et.~al (2018), the EM-REG estimator.\cite{boatman18} Instead of the mixture density of the biomarker in (\ref{eq:mix}), we could model the mixture density of the joint biomarker and study outcome by factorization:
\begin{equation*}
\begin{split}
&\lambda(b_{ij},y_{ij}|z_{ij},l_{ij-1},x_i,a_i,d_{ij};\delta,\xi,\beta_Y) \\ &= \omega(z_{ij},l_{ij-1},x_i,a_i,d_{ij};\delta)g(b_{ij}|y_{ij},z_{ij},l_{ij-1},x_i,a_i,d_{ij},c_{ij}=1;\xi)h(y_{ij}|z_{ij},l_{ij-1},x_i,a_i,d_{ij},c_{ij}=1;\beta_Y) \\ &+ (1-\omega(z_{ij},l_{ij-1},x_i,a_i,d_{ij};\delta))g(b_{ij}|y_{ij},z_{ij},l_{ij-1},x_i,a_i,d_{ij},c_{ij}=0;\xi)h(y_{ij}|z_{ij},l_{ij-1},x_i,a_i,d_{ij},c_{ij}=0;\beta_Y),
\end{split}
\end{equation*}
\sloppy where $\omega(z_{ij},l_{ij-1},x_i,a_i,d_{ij};\delta) = P(C_{ij}=1|Z_{ij}=Z_{ij},L_{ij-1}=l_{ij-1},X_i=x_i,A_i=a_i,D_{ij};\delta)$ indexed by parameter vector $\delta$. Note that both $g(b_{ij}|y_{ij},z_{ij},l_{ij-1},x_i,a_i,d_{ij},c_{ij}=1;\xi)$ and $h(y_{ij}|z_{ij},l_{ij-1},x_i,a_i,d_{ij},c_{ij}=1;\beta_Y)$ are the same densities as in (\ref{eq:mix}) and (\ref{eq:f}), respectively. We assume the same link functions for the models of $C_{ij}, B_{ij},$ and $Y_{ij}$ as before with G-computation. Again, we use the EM algorithm to derive the maximum likelihood estimates of the mixture density. We can estimate $E(Y_{iK}^*(a_i,\bar{c}_{iK}=\{1,\ldots,1\}))$ as the estimated conditional expected value of $Y_{iK}$ at the last time point:
\begin{equation*}
\frac{1}{n} \sum\limits_{i = 1}^n \hat{E}(Y_{iK}=y_{ik}|Z_{iK}=z_{iK},L_{iK-1}=l_{iK-1},X_i=x_i,A_i=a_i,C_{iK}=1;\beta_Y).
\end{equation*}
We now introduce causal estimators based off of our proposed G-computation algorithm. To illustrate the efficiency gained from implementing the full algorithm, we derive truncated G-computation estimators that ignore the two modifications we made to the algorithm. First, we have an estimator that does not use predictive mean matching and instead directly samples confounders from normal densities erroneously fit to (\ref{eq:f}). Second, we have an estimator that treats self-reported compliance as accurate by substituting in $I(D_{ij} = 1)$ for $I(C_{ij} = 1)$ in (\ref{eq:s}). Additionally, we consider a G-computation estimator with predictive mean matching where true compliance is known as in (\ref{eq:s}). Though this estimator cannot be fit in CENIC-P2, we include it for comparison to illustrate the costs of relying on estimated probabilities of compliance. We thus have 7 estimators of $E(Y_{iK}^*(a_i,\bar{c}_{iK}=\{1,\ldots,1\}))$. The ITT estimator is only included in CENIC-P2 for comparison:
\begin{enumerate}
\item The ITT estimator (only in CENIC-P2).
\item The Per Protocol estimator based off of self-reported compliance.
\item EM-REG from Boatman et.~al.
\item G-computation without predictive mean matching.
\item G-computation with self-reported compliance.
\item G-computation with true compliance (only in simulation).
\item Full G-computation (i.e., G-computation with compliance unknown and predictive mean matching).
\end{enumerate}

\subsection{Simulation Results}

For sample sizes $n = 500$ and $n = 1000$ across the three levels of $R^2$ for the confounding in the model of the study outcome, Tables (\ref{fig:sim03}--\ref{fig:sim07}) display the bias, MC SD, and MSE for each causal estimator. As $R^2$ increases across both sample sizes, the MSE of each estimator decreases largely due to the drop in MC SD. A similar observation can be made when the sample size increases across $R^2$.

Examining each estimator, the full G-computation estimator consistently returned the lowest MSE among estimators fitted to the observed data, and only had marginally higher MSE relative to G-computation with true compliance. Without predictive mean matching, G-computation returned higher MSE, though the gap closed as $R^2$ increased. Relative to the other estimators, the EM-REG estimator performed best when $R^2$ was low at 0.3. Per Protocol and G-computation with self-reported compliance had the highest MSEs, due to having the highest biases.

\newcolumntype{Z}{>{\phantom{$-$}}l}       
\newcommand\mcL[1]{\multicolumn{1}{Z}{#1}} 

\captionsetup[table]{labelfont=bf,font=normalsize,justification=raggedright,singlelinecheck=false}

\begin{table}[H]
\caption{Simulation results with $R^2 = 0.3$ for the confounding in the model of the study outcome across time points $j \in \{0,\ldots,5\}$. For both $n = 500$ and $n = 1000$, approximately 80\% of observations across simulations had self-reported compliance and thus contributed to parameter estimation.}
\begin{tabular}{llSSS} \hline
$n$ & Estimator                                                                                & \mcL{Bias}  & \mcL{MC SD} & \mcL{MSE}  \\ \hline
500       & Per Protocol                                                                             & 1.049 & 0.123 & 1.115 \\
           & EM-REG                                                                                   & 0.238 & 0.169 & 0.085 \\
           & \begin{tabular}[c]{@{}l@{}}G-computation without\\ predictive mean matching\end{tabular} & 0.187 & 0.396 & 0.191 \\           
           & \begin{tabular}[c]{@{}l@{}}G-computation with\\ self-reported compliance\end{tabular}    & 1.153 & 0.090 & 1.337 \\
           & \begin{tabular}[c]{@{}l@{}}G-computation with\\ true compliance\end{tabular}             & 0.028 & 0.100 & 0.011 \\
           & \begin{tabular}[c]{@{}l@{}}Full G-computation\end{tabular}           & 0.034 & 0.210 & 0.045 \\ \hline
1000       & Per Protocol                                                                             & 1.038 & 0.085 & 1.085 \\
           & EM-REG                                                                                   & 0.248 & 0.158 & 0.086 \\
           & \begin{tabular}[c]{@{}l@{}}G-computation without\\ predictive mean matching\end{tabular} & 0.140 & 0.313 & 0.117 \\           
           & \begin{tabular}[c]{@{}l@{}}G-computation with\\ self-reported compliance\end{tabular}    & 1.148 & 0.061 & 1.322 \\
           & \begin{tabular}[c]{@{}l@{}}G-computation with\\ true compliance\end{tabular}             & 0.027 & 0.068 & 0.005 \\
           & \begin{tabular}[c]{@{}l@{}}Full G-computation\end{tabular}           & 0.019 & 0.148 & 0.022 \\ \hline
\end{tabular}
\label{fig:sim03}
\end{table}

\begin{table}[H]
\caption{Simulation results with $R^2 = 0.5$ for the confounding in the model of the study outcome across time points $j \in \{0,\ldots,5\}$. For both $n = 500$ and $n = 1000$, approximately 80\% of observations across simulations had self-reported compliance and thus contributed to parameter estimation.}
\begin{tabular}{llSSS} \hline
$n$ & Estimator                                                                                & \mcL{Bias}  & \mcL{MC SD} & \mcL{MSE}  \\ \hline
500       & Per Protocol                                                                             & 1.032 & 0.091 & 1.073 \\
           & EM-REG                                                                                   & 0.198 & 0.106 & 0.050 \\
           & \begin{tabular}[c]{@{}l@{}}G-computation without\\ predictive mean matching\end{tabular} & 0.055 & 0.166 & 0.030 \\           
           & \begin{tabular}[c]{@{}l@{}}G-computation with\\ self-reported compliance\end{tabular}    & 1.129 & 0.069 & 1.279 \\
           & \begin{tabular}[c]{@{}l@{}}G-computation with\\ true compliance\end{tabular}             & 0.023 & 0.065 & 0.005 \\
           & \begin{tabular}[c]{@{}l@{}}Full G-computation\end{tabular}           & -0.021 & 0.097 & 0.010 \\ \hline
1000       & Per Protocol                                                                             & 1.037 & 0.065 & 1.079 \\
           & EM-REG                                                                                   & 0.189 & 0.077 & 0.042 \\
           & \begin{tabular}[c]{@{}l@{}}G-computation without\\ predictive mean matching\end{tabular} & 0.042 & 0.110 & 0.014 \\           
           & \begin{tabular}[c]{@{}l@{}}G-computation with\\ self-reported compliance\end{tabular}    & 1.131 & 0.050 & 1.282 \\
           & \begin{tabular}[c]{@{}l@{}}G-computation with\\ true compliance\end{tabular}             & 0.026 & 0.048 & 0.003 \\
           & \begin{tabular}[c]{@{}l@{}}Full G-computation\end{tabular}           & -0.022 & 0.069 & 0.005 \\ \hline
\end{tabular}
\label{fig:sim05}
\end{table}

\begin{table}[H]
\caption{Simulation results with $R^2 = 0.7$ for the confounding in the model of the study outcome across time points $j \in \{0,\ldots,5\}$. For both $n = 500$ and $n = 1000$, approximately 80\% of observations across simulations had self-reported compliance and thus contributed to parameter estimation.}
\begin{tabular}{llSSS} \hline
$n$ & Estimator                                                                                & \mcL{Bias}  & \mcL{MC SD} & \mcL{MSE}  \\ \hline
500       & Per Protocol                                                                             & 1.036 & 0.075 & 1.079 \\
           & EM-REG                                                                       & 0.177 & 0.053 & 0.034 \\
           & \begin{tabular}[c]{@{}l@{}}G-computation without\\ predictive mean matching\end{tabular} & 0.030 & 0.069 & 0.006 \\           
           & \begin{tabular}[c]{@{}l@{}}G-computation with\\ self-reported compliance\end{tabular}    & 1.125 & 0.056 & 1.268 \\
           & \begin{tabular}[c]{@{}l@{}}G-computation with\\ true compliance\end{tabular}             & 0.028 & 0.051 & 0.003 \\
           & \begin{tabular}[c]{@{}l@{}}Full G-computation\end{tabular}           & -0.030 & 0.059 & 0.004 \\ \hline
1000       & Per Protocol                                                                             & 1.031 & 0.053 & 1.066 \\
           & EM-REG                                                                                   & 0.177 & 0.037 & 0.033 \\
           & \begin{tabular}[c]{@{}l@{}}G-computation without\\ predictive mean matching\end{tabular} & 0.028 & 0.048 & 0.003 \\           
           & \begin{tabular}[c]{@{}l@{}}G-computation with\\ self-reported compliance\end{tabular}    & 1.123 & 0.043 & 1.264 \\
           & \begin{tabular}[c]{@{}l@{}}G-computation with\\ true compliance\end{tabular}             & 0.025 & 0.038 & 0.002 \\
           & \begin{tabular}[c]{@{}l@{}}Full G-computation\end{tabular}           & -0.028 & 0.042 & 0.003 \\ \hline
\end{tabular}
\label{fig:sim07}
\end{table}

\section{Application to CENIC-P2}

\subsection{Trial Design}

We applied our method to the data collected in CENIC-P2, which sought to evaluate the effect of VLNC cigarettes on average number of cigarettes smoked per day and other measures of smoking behaviors (NCT02139930).\cite{hatsukami18} Current smokers who had no intention of quitting were randomized to one of three treatment arms: VLNC cigarettes only, NNC cigarettes only, and gradual reduction from NNC cigarettes to VLNC cigarettes. During the follow-up period of 20 weeks with study visits every four weeks, participants were asked to only smoke the study cigarettes provided by the trial but could additionally smoke commercial cigarettes.

We are interested in estimating the causal effect of VLNC versus NNC cigarettes (where $a_i = \text{VLNC}$ and $a_i = \text{NNC}$, respectively) on average number of cigarettes smoked per day ($Y_{ij}$) at the end of the trial. As the NNC cigarettes have the same nicotine content as commercial cigarettes and because nicotine is thought to be primarily responsible for smoking behavior, we assumed that compliance in the NNC arm was 100\%. For time points $j \in \{0,\ldots,5\}$ corresponding to study visits at weeks $\{0,4,\ldots,20\}$, we are interested in estimating the causal effect for VLNC cigarettes at week 20: $E(Y_{i5}^*(a_i=\text{VLNC},\bar{c}_{i5}=\{1,\ldots,1\})$.

Given that the VLNC study cigarettes were not commercially available, and that CENIC-P2 enrolled current smokers who had no intention of quitting, participants that missed visits were likely smoking non-study cigarettes with normal nicotine content. Thus, we assumed that missing data indicated that participants were noncompliant.

Based on the findings of various regulatory tobacco trials,\cite{donny15,hatsukami18,smith19} for the VLNC arm we assumed that missing data indicated that participants were noncompliant. As we are only interested in estimating the causal mean among compliant participants, we can exclude missing data without introducing bias in our causal estimators.

At each study visit, participants could report any use of non-study cigarettes. We assumed that self-reports of noncompliance were accurate; therefore, observations with $D_{ij} = 0$ did not contribute to parameter estimation. For the VLNC arm of CENIC-P2, we thus have 287 of 411 (69.8\%) participants who self-reported compliance for at least one time point after baseline, where $K = 6$ time points, $j \in \{0,\ldots,5\}$, and we have 1059 observations across time points. We selected TNE as the biomarker to model, as previous research has identified TNE as a reliable biomarker for detecting noncompliance.\cite{boatman19} Due to the skewness of the distribution of TNE, and the fact that it is a concentration, we modeled it on the natural logarithm scale.

We adjusted for a number of time-varying confounders for the study outcome, $Y_{ij}$. These included both measurements of $Y_{ij}$ at earlier time points and the results of several surveys of smoking behavior. For the additional set of confounders $Z_{ij}$, we have: the Minnesota Nicotine Withdrawal Scale, the Fagerstr\"{o}m Test for Nicotine Dependence, and the Cigarette Evaluation Scale for Satisfaction. We also adjusted for a number of time-invariant confounders. For $X_i$, we have: age (by year), sex (male/female), race (Caucasian/African-American/Other), education level (less than a high school degree, high school degree only, and more than a high school degree), and average number of non-study cigarettes smoked per day at baseline.

Given that each of the six models (i.e., for compliance, the biomarker, the study outcome, and the results of the three surveys) we intend to fit conditions on numerous variables collected at either baseline, the current time point, or the previous time point, we performed variable selection on the six models. First, we defined an indicator of compliance based off the value of TNE, where the cut-off of 6.41 was determined from prior research. Specifically, based on the findings of a previous study of participants who were sequestered in a hotel and only had access to VLNC cigarettes, only 5\% of participants randomized to VLNC cigarettes are expected to have TNE above 6.41 if they were fully compliant.\cite{denlinger16} For each participant at each time point, we derived $C_{ij} = I(\text{TNE}_{ij} < 6.41)$. With compliance assumed to be known, we used the least absolute shrinkage and selection operator (LASSO) to identify variables to include for each model.\cite{santosa86,tibshirani96} For the compliance model, the LASSO did not identify a coefficient for time. Therefore, unlike the simulation, we assumed the same intercept across time points for the compliance model.

Note that in contrast to the simulation, all time-varying confounders are bounded. $Y_{ij}$ is bounded below at 0 while each survey item is bounded both above and below. We thus used predictive mean matching to generate data for all time-varying confounders.

To derive 95\% confidence intervals and standard errors for each of the estimators, we used the bootstrap.\cite{efron77} For each bootstrap sample, we resampled the same number of participants as in CENIC-P2 (i.e., 411). For each resampled participant, we included their full set of observations across time points from the observed dataset. We generated 1,000 bootstrap samples, each with $R = 10,000$ samples for estimating the causal effect in the G-computation algorithm.

\subsection{CENIC-P2 Results}

Table \ref{fig:CP2} displays the different causal estimators for the effect of VLNC vs. NNC cigarettes on average number of cigarettes smoked per day. Relative to ITT, each causal estimator returned a smaller effect of VLNC cigarettes with higher standard error. The 95\% confidence intervals for each causal estimator did not include 0. Examining each estimator, the EM-REG estimator gave the weakest effect at almost three cigarettes less than ITT. Predictive mean matching mitigated the causal effect estimate by over one cigarette per day. The full G-computation estimator returned nearly identical results to both Per Protocol and G-computation with self-reported compliance.

\FloatBarrier

\begin{sidewaystable}
\caption{CENIC-P2 results. For the ITT and Per Protocol estimators, the sample size comprises 297 and 195 participants, respectively. For all other causal estimators, the sample size comprises the 287 participants who self-reported compliance for at least one time point after baseline, where time point $j \in \{0,\ldots,5\}$ and we have 1059 observations across time points.}
\begin{tabular}{lccccc}
\hline
Estimator                                                                               & $\hat{\mu}(\text{NNC})$ & $\hat{\mu}(\text{VLNC})$ & $\hat{\mu}(\text{NNC}) - \hat{\mu}(\text{VLNC})$ & SE & 95\% CI      \\ \hline
ITT                                                                                     & 21.48 & 13.27  & 8.21           & 0.73         & (6.78, 9.60) \\ \hline
Causal Estimators                                                                       &       &        &                &              &              \\ 
\hspace{3mm}Per Protocol                                                                            & 21.48 & 14.91  & 6.57           & 0.96         & (4.61, 8.36) \\ 
\hspace{3mm}EM-REG                                                                            & 21.48 & 16.08  & 5.40           & 0.96         & (3.40, 7.17) \\ 
\hspace{3mm}\begin{tabular}[c]{@{}l@{}}G-computation without\\ predictive mean matching\end{tabular}       & 21.48 & 13.71  & 7.77           & 0.95         & (5.63, 9.42) \\ 
\hspace{3mm}\begin{tabular}[c]{@{}l@{}}G-computation with\\ self-reported compliance\end{tabular}       & 21.48 & 14.92  & 6.56           & 0.93         & (4.47, 8.31) \\ 
\hspace{3mm}\begin{tabular}[c]{@{}l@{}}Full G-computation\end{tabular}          & 21.48 & 14.97  & 6.51           & 0.92         & (4.72, 8.50) \\ \hline
\end{tabular}
\label{fig:CP2}
\end{sidewaystable}

\FloatBarrier

\section{Discussion}

For longitudinal trials with unobserved compliance, causal effects estimation can be difficult due to the potentially high number of compliance patterns. Additionally, the densities of the confounders may be difficult to parametrically model across time points. We have developed a G-computation estimator that addresses both problems. First, we simplified the longitudinal modeling structure which allowed us to calculate probabilities of compliance over time. Then, we used predictive mean matching to generate random samples of the time-varying confounders that more closely matched the empirical distributions.

Across both different levels of confounding and sample sizes, our simulation study confirms that the full G-computation estimator performs better than existing causal estimators. Both the probabilities of compliance and predictive mean matching remain important to minimizing MSE. Even when there was a single confounder fit with a semiparametric model which was not the study outcome, predictive mean matching improved estimation of the causal mean. Relative to the other estimators, full G-computation provided the largest gain in efficiency when confounding was low but not insignificant. These gains were mostly due to predictive mean matching. Moreover, with low confounding, the conditional confounder distributions are less concentrated around the mean. Predictive mean matching was better able to reproduce the spread of the empirical distributions.

In our application to the data collected in CENIC-P2, each causal estimator returned a significant if reduced effect of VLNC cigarettes relative to ITT. These findings confirm the original conclusion of the trial, which is that VLNC cigarettes can reduce cigarette consumption. Interestingly, the full G-computation estimator returned similar results to G-computation with self-reported compliance, implying that self-reports may have been accurate for CENIC-P2. Additionally, both estimators returned similar results to Per Protocol, implying that there may not have been much confounding for the study outcome. Predictive mean matching, however, was necessary for properly modeling the confounders.

The primary contribution of our method is that it provides a framework for modeling multiple longitudinal variables which standard regression-based methods may be ill-suited to handle. Provided that there is a biomarker strongly associated with compliance, probabilities of compliance can be inserted into the G-computation algorithm when compliance is unknown. When randomly sampling in G-computation, predictive mean matching can loosen the parametric assumptions of the confounder densities. As demonstrated in the simulation, the causal estimator with these two modifications is asymptotically unbiased.

The limitations of our method mainly relate to the complexities of dealing with longitudinal data and numerous confounders. To simplify the longitudinal modeling structure, we assumed that observations only depended on data collected at the previous time point or baseline. This assumption may not hold when there is autocorrelation between observations across time points. In such a scenario, the score equations in (\ref{eq:full}) may be more appropriate. Longitudinal compliance probabilities could be substituted for the longitudinal indicator of compliance in (\ref{eq:full}). However, this would require the fit of a longitudinal mixture density (e.g, with mixed effects models), which would be computationally intensive with many covariates. 

Additionally, while predictive mean matching can improve randomly sampling from the confounder densities, errors may accumulate with large numbers of confounders. Variable selection may also be difficult without a biomarker that has a strong association with compliance. In the application to CENIC-P2, we were able to implement the LASSO by defining a compliance indicator based off of TNE, which has an AUC of 0.99 with compliance.\cite{boatman19} With a biomarker that has a weaker association with compliance, the LASSO may not be as reliable.

In secondary analyses of longitudinal trials, the G-computation algorithm can estimate the causal effect for potentially large numbers of treatment combinations. When compliance and hence the treatment received are unknown, we were able to simplify the G-computation algorithm and insert probabilities of compliance to estimate the causal effect for a single compliance pattern of interest, namely, full compliance with the trial. Furthermore, the G-computation algorithm depends on correctly specifying the models of the confounders, of which there may be many. Randomly sampling from multiple misspecified models can lead to bias and variability in estimates of the causal effect. Our analysis confirms that predictive mean matching can generate confounder data that more closely mirror the true distributions of the confounders. Overall, we have demonstrated that under certain assumptions, the G-computation algorithm can be implemented when compliance is unknown. Predictive mean matching should be considered when the densities of the confounders substantially deviate from parametric densities.

\section{Acknowledgements}

The author(s) disclosed receipt of the following financial support
for the research, authorship, and/or publication of this
article: This study was funded by the National Heart, Lung,
and Blood Institute (award number T32HL129956), the
National Cancer Institute (award numbers R01CA214825
and R01CA225190), the National Institute on Drug Abuse
(award numbers R01DA046320, R03DA041870, and U54-
DA031659) and National Center for Advancing Translational
Science (award number UL1TR002494). The content is solely
the responsibility of the authors and does not necessarily represent
the official views of the National Institutes of Health
or Food and Drug Administration.

\bibliography{citations}

\clearpage

\setcounter{page}{1}

\section{Supplementary Material}

\subsection{Proof that the Expected Value of (\ref{eq:s}) is Equal to the Expected Value of (\ref{eq:wlog})} \label{ie}

The expected value of (\ref{eq:s}) is given as:
\begin{equation*}
\begin{split}
&\sum \limits_{j=1}^K \sum\limits_{i=1}^n E[I(C_{ij}=1) \frac{\partial}{\partial \beta} \log f(L_{ij}|L_{ij-1},X_i,A_i=a_i,C_{ij}=1;\beta)].
\intertext{By iterated expectation:}
&=\sum \limits_{j=1}^K \sum\limits_{i=1}^n E[E[I(C_{ij}=1) \frac{\partial}{\partial \beta} \log f(L_{ij}|L_{ij-1},X_i,A_i=a_i,C_{ij}=1;\beta)|L_{ij},L_{ij-1},B_{ij},X_i,D_{ij},A_i=a_i]]
\\&=\sum \limits_{j=1}^K \sum\limits_{i=1}^n E[E[I(C_{ij}=1)|L_{ij},L_{ij-1},B_{ij},X_i,D_{ij},A_i=a_i] \frac{\partial}{\partial \beta} \log f(L_{ij}|L_{ij-1},X_i,A_i=a_i,C_{ij}=1;\beta)]
\\&=\sum \limits_{j=1}^K \sum\limits_{i=1}^n E[P(C_{ij}=1|L_{ij},L_{ij-1},B_{ij},X_i,D_{ij},A_i=a_i) \log \frac{\partial}{\partial \beta} f(L_{ij}|L_{ij-1},X_i,A_i=a_i,C_{ij}=1;\beta)],
\end{split}
\end{equation*}
where the last expression is the expected value of (\ref{eq:wlog}).

\begin{landscape}

\subsection{EM Algorithm Step Updates} \label{ca}

We delineate the EM algorithm step updates to derive the maximum likelihood estimates of the vector of parameters for the mixture density below:
\begin{equation*}
\begin{split}
&\psi(b_{ij}|l_{ij},l_{ij-1},x_i,a_i,d_{ij};\alpha,\xi) \\&= \rho(l_{ij},l_{ij-1},x_i,a_i,d_{ij};\alpha)g(b_{ij}|l_{ij},l_{ij-1},x_i,a_i,d_{ij},c_{ij}=1,\xi) \\ &+ (1-\rho(l_{ij},l_{ij-1},x_i,a_i,d_{ij};\alpha))g(b_{ij}|l_{ij},l_{ij-1},x_i,a_i,d_{ij},c_{ij}=0;\xi).
\end{split}
\end{equation*}
where $\rho(l_{ij},l_{ij-1},x_i,a_i,d_{ij};\alpha) = P(C_{ij}=1|L_{ij}=l_{ij},L_{ij-1}=l_{ij-1},X_i=x_i,A_i=a_i,D_{ij};\alpha)$ indexed by parameter vector $\alpha$ and $g$ is the conditional density of $B_{ij}$ indexed by parameter vector $\xi$. The subscripts denote the $i$th participant at the $j$th time point for $i \in \{1,\ldots,n\}$ and $j \in \{1,\ldots,K\}$.

Let $\boldsymbol{\theta} = \{\alpha,\xi\}$ denote the vector of parameters. Random vectors are denoted with bold font, where the subscript denotes the time point. For example, for time point 1, the set of all biomarker values is denoted as $\boldsymbol{B_1} = (B_{11},\ldots,B_{n1})^T$. The complete data log likelihood for $\boldsymbol{\theta}$ is: \footnotesize
\begin{equation*}
\begin{split}
&\log(\boldsymbol{\theta}|\boldsymbol{B_1, \ldots, B_K, C_1, \ldots, C_K, L_0, \ldots, L_K, X, A, D_1, \ldots, D_K}) \\
&\begin{split}
= \sum_{i=1}^n \sum_{j=1}^{K} & I(C_{ij} = 1) \log g(b_{ij}|L_{ij},L_{ij-1},X_i,A_i,D_{ij},C_{ij}=1;\xi) + (1 - I(C_{ij} = 1)) \log g(b_{ij}|L_{ij},L_{ij-1},X_i,A_i,D_{ij},C_{ij}=0;\xi) \\
&+ I(C_{ij}=1) \log\rho(L_{ij},L_{ij-1},X_i,A_i,D_{ij};\alpha) + (1-I(C_{ij}=1)) \log(1-\rho(L_{ij},L_{ij-1},X_i,A_i,D_{ij};\alpha)),
\end{split}
\end{split}
\end{equation*} \normalsize
where $\log$ denotes the natural logarithm and $I$ denotes the indicator function.
The conditional expectation of the E-step is given by: \footnotesize
\begin{equation*}
\begin{split}
&Q(\boldsymbol{\theta},\boldsymbol{\theta}^{(v)},\boldsymbol{B_1, \ldots, B_K, C_1, \ldots, C_K, L_0, \ldots, L_K, X, A, D_1, \ldots, D_K})\\
&=E_{\boldsymbol{\theta}^{(v)}}[\log(\boldsymbol{\theta}|\boldsymbol{B_1, \ldots, B_K, C_1, \ldots, C_K, L_0, \ldots, L_K, X, A, D_1, \ldots, D_K})|\boldsymbol{B_1, \ldots, B_K}] \\
&\begin{split}
= \sum_{i=1}^n \sum_{j=1}^{K} & w_{ij}^{(v)} \log g(b_{ij}|L_{ij},L_{ij-1},X_i,A_i,D_{ij},C_{ij}=1;\xi) + (1 - w_{ij}^{(v)}) \log g(b_{ij}|L_{ij},L_{ij-1},X_i,A_i,D_{ij},C_{ij}=0;\xi) \\
&+ w_{ij}^{(v)} \log\rho(L_{ij},L_{ij-1},X_i,A_i,D_{ij};\alpha) + (1-w_{ij}^{(v)}) \log(1-\rho(L_{ij},L_{ij-1},X_i,A_i,D_{ij};\alpha)),
\end{split}
\end{split} 
\end{equation*} \normalsize
where:
\begin{equation*}
\begin{split}
&w_{ij}^{(v)} = E_{\boldsymbol{\theta}^{(v)}}(I(C_{ij}=1)|L_{ij},L_{ij-1},X_i,A_i,D_{ij}) \\
&= \frac{\rho(L_{ij},L_{ij-1},X_i,A_i,D_{ij};\alpha^{(v)})g(B_{ij}|L_{ij},L_{ij-1},X_i,A_i,D_{ij},C_{ij}=1;\xi^{(v)})}{\psi(B_{ij}|L_{ij},L_{ij-1},X_i,A_i,D_{ij};\alpha^{(v)},\xi^{(v)})},
\end{split}
\end{equation*}
where the denominator is equal to:
\begin{equation*}
\begin{split}
&\psi(B_{ij}|L_{ij},L_{ij-1},X_i,A_i,D_{ij};\alpha^{(v)},\xi^{(v)})\\&=
\rho(L_{ij},L_{ij-1},X_i,A_i,D_{ij};\alpha^{(v)})g(B_{ij}|L_{ij},L_{ij-1},X_i,A_i,D_{ij},C_{ij}=1;\xi^{(v)})\\&+ (1-\rho(L_{ij},L_{ij-1},X_i,A_i,D_{ij};\alpha^{(v)}))g(B_{ij}|L_{ij},L_{ij-1},X_i,A_i,D_{ij},C_{ij}=0;\xi^{(v)}).
\end{split}
\end{equation*}
In weight $w_{ij}^{(v)}$, the $(v)$ denotes the $v$th iteration of the EM algorithm. The M-step update is provided by the fit of the logistic and linear regression models with weights $w_{ij}^{(v)}$ added to the respective log likelihoods.

\end{landscape}

\end{document}